# Left of Fab: Securing Design and Collaboration in the Semiconductor Value Chain

John C. Hoag
The University of Akron, hoagjc@uakron.edu

*Abstract* - **The purpose of this paper is to fill a gap in the general understanding - and academic scrutiny - of current and emerging workflows for designing and fabricating integrated circuits. The approach is to compare the IC design workflow with that for printed circuit boards, then to discern a classification for threats. The need to define and secure workflows is amplified by both U.S. investment in the semiconductor manufacturing and market forces affecting GPU production for AI applications. The origin of this knowledge gap can be the proprietary nature of solution spaces, but it can be the lack of demand for teaching and learning for engineers and technicians in this domain. This paper presents a framework for understanding the security of design workflows in a vendor- and tool-agnostic way.**

### INTRODUCTION

In his study of resiliency, Woods posits that the origins of later system failures are often present at the time of design, termed latent failures. [1] Later cyber exploits represent vulnerabilities that are latent artifacts of design flaws. While some exploits are simply opportunistic – consider theft or vandalism – our goal is to make design workflows less opaque, revealing processes and interface such that they can be made more resilient. These processes increasingly resemble software development, leveraged on re-use of libraries – literally blocks of intellectual property (IP) – with the intent of shortening time-to-market, reducing errors, and increasing yield. In short, the goal is to correct mistakes "on the drawing board" rather than during prototyping or production.

The title of this paper, consistent with the theme of cybersecurity, relates to military idiom and book entitled, <u>Left of Boom</u>, which espoused a preventative approach to global hostility.[2] Our intent is to be similarly proactive in identifying the pre-production design period prior to, or "left of fabrication." The title tacitly acknowledges presence of adversaries, whether business competitors, or other national interests' intent on exploiting semiconductor vulnerabilities.

To decouple semiconductor design from fabrication has been proven to be model for industry success; at one level, this "fabless" process permits a firm with sufficient intellectual property (IP) provide the entire detailed design for handoff to a contract manufacturer. But at a practical level, a design can consist of IP blocks licensed from several sources with the intent to interconnect and package on a custom basis, which can include production at scale.

This paper aligns elements and attributes of design with cybersecurity principles, mapping isolated topics into taxonomy and ontology of cybersecurity. This paper strives to be neutral, forbearing from all references to vendors, products, and even known faults.

### CYBERSECURITY OVERVIEW

A first-principles approach to cybersecurity maps each threat to a violation of confidentiality, integrity, availability, or nonrepudiation, commonly terms the "CIA Triad." Regarding proprietary IP, in all or part, we may equate confidentiality with privacy and secrecy, integrity with corruption, availability with denial of service (to be defined), and nonrepudiation of responsibility (for some transaction or interaction).

Confidentiality. Clearly, theft of design IP is highly problematic, irreversible, with few means of recourse. This affects IP blocks in the design phase as well as embedded IP at point of manufacture. Theft of IP can lead to counterfeit devices and entire products. Legal means to license exist, and technical means to authenticate and authorize usage are feasible. Very limited means of attestation such as device fingerprinting are feasible, though serialization at the device level is not. Privacy and secrecy are related concepts.

Integrity maps to the unaltered, attested version of a product both in its digital and physical forms.

Availability is often defined as its antonym, denial of service, whose scope can be as granular as a less-functional operational mode or interruptions and delays of over fulfillment. To correct design and manufacturing flaws would be a new form of denial of service for semiconductors.

Nonrepudiation pertains to a transaction between parties, here relating to design or fulfillment, where responsibility is assured.

A concluding section of this paper discusses "right of fab," specifically destructive and nondestructive examination of actual devices and their functionality. This section does not examine consequences in the physical setting of cyber-physical systems, where harmful operations modes can be triggered; this topic is beyond the scope of this paper.

### SCOPE OF 'LEFT OF FAB' FOR PRINTED CIRCUIT BOARDS.

One starting point for describing Left of Fab is to compare and contrast the semiconductor ecosystem with printed circuit boards, relating both design and manufacture processes. Note first that the feature size in semiconductors is measured in nanometers, commonly 30 nm but with emerging processes o for 3 nm features; in contrast, multilayer circuit board features are as small as 0.3 millimeters, a scale difference on the order of a million times. Notional layering exists in the production of each, with a similar foundational production cycle based on lithography and deposition of patterns, in that the deliverable of the both design processes is termed tapeout – indeed a "throwback" to the manual process of masking early single-sided circuit boards at full scale.

The logic board in a modern smartphone, for instance, will have on the order of ten layers, some of which extend functional traces in three dimensions, while others will be for power delivery and signal isolation. Production of both PCBs and, in fact, complete assembly at the board level is a global enterprise with an established standard for sharing design documents – consisting of vectorized images per layer and details for circuit "via" and drill positions. [3]

Tools for computer aided design support digital, analog, and mixed product development and subsume schematic capture, circuit board layout, and related simulations for testing. At this level of design, inputs include libraries of component products, and outputs include the bill-of-material. To the extent that functional requirements and constraints are known, designs can be exercised through simulation and thus validated, mitigating risk.

Note that downstream risks are nontrivial at the board level, subsuming counterfeit (or compromised) parts as well as integrity of the as-built design. At a higher level are concerns over embedded firmware or other unexpected modes of operation. These supply chain compromises are a known typology within the security community but are rarely recognized in the open literature. [4]

### SCOPE OF 'LEFT OF FAB' FOR INTEGRATED CIRCUITS

The PCB workflow is efficient and mature, to the point of open-source tools and a global supply chain – with the expectation of high yield. The integrated circuit workflow is different – and not simply a matter of "macro vs micro" analysis. To be specific, the scale comparison is milli- versus nano- where the considerations are millimeter level board layer assembly, and at a dimension reduction of $10^{-6}$ single nanometer scale construction of single part, in *extermis, one* atom at a time.

The integrated circuit workflow involves a complicated and iterative design process whose work product consists of the design of circuits at the transistor level but instructions for a multi-step assembly process one layer of materials at a time. Built upon a chemically doped, epitaxially-grown substrate, each layer joins in a repetitive process of photolithography, etching, material deposition, and cleaning; within this growing two-dimensional structure are transistor, gate, and register-level interconnections – as well as entire serialized networks – along with distribution of power and timing for synchronization. From the "front-end" to the 'back-end" of this process includes external interfaces for the package itself. The reader is directed to the Nature article and the Lienig text for insights into the contours and mechanisms of floorplanning and tapeout. [5] Similar to Gerber plots for outsourced PCB manufacture, standards-based interface formats serve the handoff from designer to the fab. [6]

The inputs to this process can be custom ideation, but, as in software development, there is value in reusable libraries termed Intellectual Property (IP) which can be incorporated in blocks. Such IP blocks include processors, memory, communications subsystems, and sensors; moreover, these designs many serve digital, analog, or mixed-signal applications. These blocks will "meet" at the design palette and be joined, subject to validation, in the fabrication process into a single die to be packaged.

As in many domains, this approach to modular design emphasizes simulation to capture and correct issues specifically with synchronization, power draw, and thermal effects, as well as invocation of design heuristics that ensure functionality and manufacturability. As these libraries of so-called "hard IP" and "soft IP" converge into a design, interconnection and interoperation merit attention. To create and curate IP is within our scope.

The ambient U.S. discussion [6] of semiconductor production as a matter of national security, competition to develop engines for large-scale artificial intelligence by information service providers and other "fabless" entrants is profound, both in datacenter and edge cloud settings.[7] This is significant "Left of Fab" in that these firms, to characterize broadly, are new to the semiconductor industry and, given their primary line of business is something else, are not specifically exposed to cyber risk of their new devices. Handset vendors have substantially more experience fielding Systems-on-Chips while cloud providers are new entrants.

Among the archetypal "CIA Triad" of cyber vulnerabilities design concerns can be cast in several ways

- Confidentiality, as in theft or loss of IP
- Integrity, as in untraceable modifications to IP
- Availability, as in "denial of market" due reworking designs, flawed finished goods, or low yields.
- Non-repudiation, though remote, of lack of traceability in the design flow

A more interesting speculative case is corruption of the verification/validation test cases, which could lead to "denial of" cases above; in the next section, we mention corruption or loss of anti-side-channel countermeasures, which could lead to devices with exploitable features.

In a very recent development, the March 2024 National Strategy on Microelectronics Research under the seal of Office of the U.S. President, calls for adoption of "secure cloud-based solutions" based on interoperable standards-based Product Development Kits in the domain of Electronic Design Automation. [9] The intent of this activity is to accelerate semiconductor product development, subject to rigorous protection of IP and respect for export controls. This activity is expected to expand the defense industrial base, characterized as low volume, high mix products with special interest in use cases like radio frequency, with digital twins. Industry leaders expect a role for A.I. in future design.[10]

This document does not expand the scope of our interests in semiconductor design cybersecurity, only its scale and timeliness. It does infer new roles for the U.S. government in marshalling collaborative design, as well as underscore the need to standardize material and process definitions.

Given that the overall mission for this project is to elevate systems-level knowledge of workflows through fabrication, our challenge is to document and propagate knowledge.

## FOR CLOSURE: 'RIGHT OF FAB'

For sake of completeness, this paper briefly recognizes an entire genre of vulnerabilities once devices are fielded. Manufacturers must expect all devices to be subject to batteries of *in situ* and destructive tests that may reveal the contents of memory or the operation of algorithms, most sensitive when those contents are encryption methods, or the keys created. Given that expectation, design flows can be modified to avoid such exploits; control over the test cases can be considered sensitive metadata, also highly valued to a motivated attacker.

The CVE (cf. Common Enumeration of Vulnerabilities) is an authoritative U.S. curation of disclosed techniques, culminating in the SPECTRE class of vulnerabilities. In general, all RAM contents linger and are accessible briefly; processor cache contents, needed to provide the perception of greater performance, are susceptible to inspection, both locally, and perhaps remotely. While this is synchronous and in situ, asynchronous techniques serve to reveal nonvolatile memory contents of data and code when suitably exercised.

This is an active domain of research both by malevolent actors and IP holders, perhaps a balance between the pursuit of [cached] performance and the privacy of memory.

## CONCLUSION

This paper has provided a vendor-, tool- and manufacturer-agnostic overview of semiconductor design workflows with a perspective on latent flaws along the classic dimensions of confidentiality, integrity, availability, and nonrepudiation. The design process is perceived as opaque, not due to proprietary interests but more due to gaps in the literature.

The narrow mission of this project is to help students enter the semiconductor workforce, and while many will join as manufacturing associates, others may examine the value chain and elect to participate in other roles, particularly design. Cyber is a thread across all design domains, and the integrated nature of IP is certain to attract malevolent, motivated actors – to steal compromise products and data.

## REFERENCES

[1] Woods, D.D., et al, Behind Human Error, 2e, Taylor and Francis, 2017.
[2] Laux, D., Left of Boom, St. Martin's, 2016
[3] https://www.ucamco.com/en/gerber, accessed April 23, 2024.
[4] https://attack.mitre.org/techniques/T0862/ accessed April 23, 2024.
[5] Inter alia: Semiconductor Processing, P.A. Rockwell, 2023; Semiconductor Microchips and Fabrication by Y. Lian, IEEE/Wiley, 2023; and specifically, "A Graph Placement Methodology for Fast Chip Design" in Nature 594, 2021; Leinig and Scheible, Fundamentals of Layout Design for Electronic Circuits, Springer, 2020.
[6] Calma's 1978 Graphic Design System interface, GDS II, is not maintained by its author, and, though not deprecated, has been surpassed by the OASIS format, an industry standard.
[7] Miller, Chris, Chip War: The Fight for the World's Most Critical Technology, Simon and Shuster, 2023
[8] Savitz, Eric J., "Nvidia Won AI's First Round. Now the Competition Is Heating Up," Barrons, April 19, 2024.
[9] "National Strategy on Microelectronics Research," Office of the U.S. President, March 2024.
[10] "How will AI Affect the Semiconductor Industry? Industry veterans consider a future 'AI Wonderland,'" IEEE Spectrum 11/23.

## AUTHOR INFORMATION

John C. Hoag, Ph.D., IEEE Senior Member, is the founding Managing Director of the 5G Broadband & Connectivity Center at Ohio State and is on the faculty at Akron University.